\documentclass[conference]{IEEEtran}
\IEEEoverridecommandlockouts
\usepackage{cite}
\usepackage{amsmath,amssymb,amsfonts}
\usepackage{algorithmic}
\usepackage{graphicx}
\usepackage{textcomp}
\usepackage{xcolor}
\usepackage{algorithmic}
\usepackage{url}
\usepackage{lipsum}
\usepackage{multirow}
\usepackage[normalem]{ulem}
\usepackage{bbding}
\usepackage{pifont}
\usepackage{wasysym}
\usepackage{bbm}
\usepackage{dsfont}
\usepackage[ruled,lined,boxed]{algorithm2e}
\usepackage{mathtools, nccmath}
\usepackage{amsthm}
\usepackage{geometry}
\DeclarePairedDelimiter{\nint}\lfloor\rceil
\DeclareMathAlphabet{\mathbcal}{OMS}{cmsy}{b}{n}
\def\BibTeX{{\rm B\kern-.05em{\sc i\kern-.025em b}\kern-.08em
		T\kern-.1667em\lower.7ex\hbox{E}\kern-.125emX}}
\newtheorem{remark}{Remark}
\renewcommand{\baselinestretch}{0.96}

\begin{document}
	
\title{DPSS-based Codebook Design for Near-Field XL-MIMO Channel Estimation\\
}
\author{
	\IEEEauthorblockN{
		Shicong~Liu\IEEEauthorrefmark{1},  Xianghao~Yu\IEEEauthorrefmark{1}, Zhen~Gao\IEEEauthorrefmark{2}, and Derrick~Wing~Kwan~Ng\IEEEauthorrefmark{3}
	}
	\IEEEauthorblockA{\IEEEauthorrefmark{1}Department of Electrical Engineering, City University of Hong Kong, Hong Kong}
	\IEEEauthorblockA{\IEEEauthorrefmark{2}Advanced Research Institute of Multidisciplinary Science, Beijing Institute of Technology, Beijing, China}
	\IEEEauthorblockA{\IEEEauthorrefmark{3}School of Electrical Engineering and Telecommunications, University of New South Wales, Sydney, Australia}
	
	Email: scliu@ieee.org, alex.yu@cityu.edu.hk, gaozhen16@bit.edu.cn, w.k.ng@unsw.edu.au
	\vspace{-4mm}
}

\maketitle

\begin{abstract}
	Future sixth-generation (6G) systems are expected to leverage extremely large-scale multiple-input multiple-output (XL-MIMO) technology, which significantly expands the range of the near-field region. While accurate channel estimation is essential for beamforming and data detection, the unique characteristics of near-field channels pose additional challenges to the effective acquisition of channel state information. In this paper, we propose a novel codebook design, which allows efficient near-field channel estimation with significantly reduced codebook size. Specifically, we consider the eigen-problem based on the near-field electromagnetic wave transmission model. Moreover, we derive the general form of the eigenvectors associated with the near-field channel matrix, revealing their noteworthy connection to the discrete prolate spheroidal sequence (DPSS). Based on the proposed near-field codebook design, we further introduce a two-step channel estimation scheme. Simulation results demonstrate that the proposed codebook design not only achieves superior sparsification performance of near-field channels with a lower leakage effect, but also significantly improves the accuracy in compressive sensing channel estimation.
\end{abstract}

	
\section{Introduction}
	
{T}{he} development of massive multiple-input multiple-output (MIMO) systems has spurred a vision to reshape and control transmission environments of electromagnetic waves, leading to the emergence of advanced technologies such as cell-free massive MIMO and reconfigurable intelligent surfaces (RIS) that enhance service coverage and eliminate dead zones in wireless networks\cite{WC21Yu,TWC17Ngo}. Particularly, for centralized large-scale antenna array deployment strategies, like RIS and extremely large-scale MIMO (XL-MIMO)\cite{WC23Wang}, their vast apertures significantly expand the boundaries of the near-field region~\cite{JSAC20Dardari}. In practice, mobile devices within the near-field region can achieve higher transmission rates, which, however, requires accurate channel state information. Unfortunately, the proliferation of antennas and distinctive properties of near-field channels introduce additional hurdles in channel estimation~(CE).

In the literature, compressive sensing (CS)-based techniques have been proposed to reduce the required excessive training overhead in CE by exploiting the intrinsic sparsity of channel matrices\cite{TIT06Donoho}. In fact, the performance of such algorithms highly depends on the codebooks that match the channel model. However, the commonly-adopted codebooks in the far-field region, e.g., discrete Fourier transform (DFT) codebook, show a severe mismatch with the near-field spherical wave transmission model, which results in an energy leakage effect in sparse representation, thereby significantly undermining the performance of CS-based algorithms. On the other hand, although the spherical wave codebook\cite{WCL19Han} matches the near-field transmission model, the columns within the codebook matrix are not mutually orthogonal, which may further cause performance degradation and jeopardize the convergence of the algorithms. Besides, the two spatial degrees of freedom (DoFs), i.e., distance and angle, in the spherical codebook result in increased storage requirements and computational complexity for codebook matching. 

As a remedy, a polar-domain sampling scheme for the spherical wave codebook was proposed\cite{TCOM22Cui}. The scheme leverages the inverse proportionality between the mutual correlation of spherical wave steering vectors and distance to significantly reduce the codebook size. Later on, a hierarchical near-field codebook was proposed, where the upper-layer codebooks are exploited for target location search while the lower-layer ones are adopted to achieve the highest beam gain around the steering points\cite{WCL22Chen}. However, the aforementioned studies are essentially refinements of the conventional spherical wave codebook, which fail to address the high mutual correlation issue among codewords. An alternative codebook design was recently presented~\cite{TWC23Shi}, which utilized the spatial-chirp beam to reduce training overhead. Also, dictionary learning was exploited in codebook design~\cite{TCOM23Zhang}, which iteratively updated the codebook and reconstructed the channel matrix. Nevertheless, the strict orthogonality among codewords still cannot be ensured and a fine-tune procedure is required for different application scenarios. Hence, designing a codebook that is not only small in size but also column-wise orthogonal remains an open problem.
	
In this paper, we address the mismatch between the DFT vectors and the spherical wave transmission model, and also tackle the non-orthogonality associated with the conventional spherical wave codebook. Specifically, we propose a lightweight yet effective codebook by exploring the eigenvalue-decomposition (EVD) of the near-field channel matrix and reveal that the corresponding eigenvectors admit the form of discrete prolate spheroidal sequences (DPSS). By constructing the codebook exploiting these orthogonal vectors, we inherently avoid an oversized codebook and ensure mutual orthogonality among the codewords. Furthermore, we propose a two-step CE scheme for near-field XL-MIMO and evaluate the performance through simulations. Numerical results demonstrate that the proposed CE scheme with the novel DPSS-based codebook achieves a significant improvement in channel sparsification, thereby contributing to higher accuracy in near-field CE compared to the DFT and spherical codebooks. More importantly, the required size of the proposed DPSS-based codebook is substantially smaller than the conventional DFT and spherical wave codebooks, which leads to less stringent storage requirements.

{\it Notations}: We use normal-face letters to denote scalars and lowercase (uppercase) boldface letters to denote column vectors (matrices). The $k$-th row vector and the $m$-th column vector of matrix ${\bf H}\in\mathbb{C}^{K\times M}$ are denoted as ${\bf H}[{k,:}]$ and ${\bf H}[{:,m}]$, respectively. $\{{\bf H}_n\}_{n=1}^N$ denotes a matrix set with the cardinality of $N$. The superscripts $(\cdot)^{T}$, $(\cdot)^{\rm *}$, $(\cdot)^{H}$, and $(\cdot)^{\dagger}$ represent the transpose, conjugate, conjugate transpose, and pseudo-inverse operators, respectively. $\mathcal{CN}(\mu,\sigma^2)$ denotes the complex Gaussian distribution with mean $\mu$ and standard deviation $\sigma$, and $\mathbb{E}[\cdot]$ denotes the statistical expectation operator. The $0$-norm of a vector $\Vert\cdot\Vert_0$ counts the number of its non-zero elements. The imaginary unit is represented as $j$ such that $j^2=-1$.
	
\section{System Model}
\begin{figure}[t]
	\centering
	\includegraphics[width=0.25\textwidth]{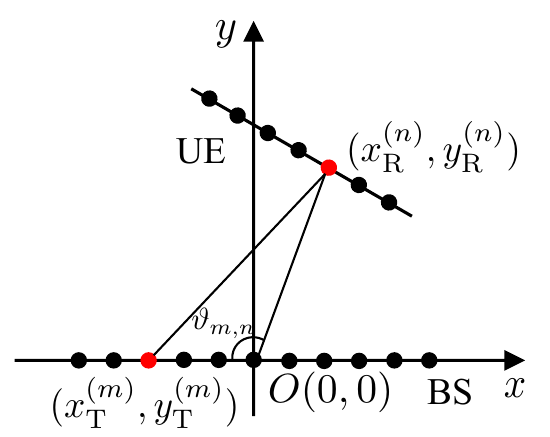}
	\caption{The considered near-field transmission scenario. The coordinates of the $m$-th element in the transmit antenna array and the $n$-th element in the receive antenna array are $(x_{\rm T}^{(m)},y_{\rm T}^{(m)})$ and $(x_{\rm R}^{(n)},y_{\rm R}^{(n)})$, respectively.}
	\label{fig:sysmodel}
\end{figure}

Consider a user equipment (UE) array\footnote{It can be extended to multi-user scenarios by assigning orthogonal pilots for different UEs.} communicates with a base station (BS) equipped with an XL-MIMO array in its near-field region. The generated electric field ${\bf E}({\bf r}_{\rm R})$ at the UE can be expressed by the integral of the spatial impulse response ${\bf G}({\bf r}_{\rm T},{\bf r}_{\rm R})$ with a current source ${\bf J}({\bf r}_{\rm T})$ at the BS as~\cite{JSAC20Dardari}
\begin{equation}
	{\bf E}({\bf r}_{\rm R}) = \int_{{\mathcal{S}_{\rm T}}} {\mathbf{G}}\left({\bf r}_{\rm T},{\bf r}_{\rm R}\right) {\bf J}({\bf r}_{\rm T})~{\rm d}{\bf r}_{\rm T},
\end{equation}
where ${\bf r}_{\rm T}=(x_{\rm T},y_{\rm T})$ and ${\bf r}_{\rm R}=(x_{\rm R},y_{\rm R})$ denote the coordinates of the transmitter and receiver, respectively, and ${\mathcal{S}_{\rm T}}$ denotes the transmit aperture. The impulse response ${\bf G}\left({\bf r}_{\rm T},{\bf r}_{\rm R}\right)$ can be derived in dyadic form~\cite{TIT05Poon} as
\begin{align}
	{\mathbf{G}}\left({\bf r}_{\rm T},{\bf r}_{\rm R}\right) = {} & {} \frac{j\kappa Z_0 e^{j\kappa \Vert {\bf r} \Vert}}{4\pi \Vert {\bf r} \Vert}\left[ \left( {\bf I}-\hat{\bf r}\hat{\bf r}^H \right)+ \frac{j}{\kappa\Vert {\bf r} \Vert}\left( {\bf I}-3\hat{\bf r}\hat{\bf r}^H \right)\right.\notag\\
	{} & {} \left. -\frac{1}{(\kappa \Vert {\bf r} \Vert)^2}\left( {\bf I}-3\hat{\bf r}\hat{\bf r}^H \right) \right]\label{green}\\
	\simeq {} & {} \varphi_0 \frac{e^{-j\kappa \Vert {\bf r}\Vert}}{\Vert{\bf r}\Vert} \left( {\bf I}-\hat{\bf r}\hat{\bf r}^H \right),\notag
\end{align}
where ${\bf I}$ denotes the identity matrix, $\varphi_0 = j\kappa Z_0/(4\pi)$, $\kappa = 2\pi/\lambda$ is the wavenumber, and $Z_0\approx 376.73 \:\Omega$ is the intrinsic impedance of free space. ${\bf r} = {\bf r}_{\rm R}-{\bf r}_{\rm T}$ and $\hat{\bf r}={\bf r}/\Vert {\bf r}\Vert$ denotes the direction of $\bf r$ with unit length. For uni-polarized antennas, the impulse response reduces to the scalar form as
\begin{equation}
	{g}({\bf r}_{\rm T},{\bf r}_{\rm R}) = \varphi_0 {e^{-j\kappa \Vert {\bf r}\Vert}}/{\Vert{\bf r}\Vert}.
	\label{eq:sv}
\end{equation}
~~Consider that both the BS and UE are equipped with uniform linear arrays (ULA)\footnote{We consider ULA here for brevity, while it can be extended to other antenna geometries. For example, it can be extended to uniform planar array (UPA) {by applying Kronecker products to steering vectors in~\eqref{eq:LoS}}.}, the near-field communication scenario is then shown in Fig.~\ref{fig:sysmodel}. For the $m$-th ($1\leq m\leq N_{\rm T}$) antenna element in the transmit array, the downlink line-of-sight (LoS) wireless channel can be modeled as 
\begin{equation}
	\begin{aligned}
		\mathbf{H}_{\rm LoS}[:,m] &= {\mathbf g}_{\rm R}({\bf r}_{\rm T}^{(m)})\\&= \left[ \tilde{g}({\bf r}_{\rm T}^{(m)},{\bf r}_{\rm R}^{(1)}),\cdots,\tilde{g}({\bf r}_{\rm T}^{(m)},{\bf r}_{\rm R}^{(N_{\rm R})}) \right]^{T},
	\end{aligned}
	\label{eq:LoS}
\end{equation}
where $N_{\rm T}$ and $N_{\rm R}$ denote the numbers of antennas at the BS and UE, respectively, and $\tilde{g}(\cdot) = g(\cdot)/\varphi_0$ is the normalized impulse response. 
Considering Rician fading, the overall downlink channel matrix can be modeled as
\begin{equation}
	{\bf H} = \sqrt{\frac{K}{1+K}}\mathbf{H}_{\rm LoS} + \sqrt{\frac{1}{1+K}}\mathbf{H}_{\rm NLoS},
	\label{eq:channelmodel}
\end{equation}
where $K\geq 0$ denotes the Rician factor. The non-line-of-sight (NLoS) channel components satisfy ${\bf H}_{\rm NLoS}[n,m]\sim\mathcal{CN}(0,\sigma^2)$, $\forall 1\leq n\leq N_{\rm R},~1\leq m\leq N_{\rm T}$ with $\sigma^2 = 1/(N_{\rm T}N_{\rm R})$.

Since XL-MIMO arrays are deployed at both the UE and BS, hybrid analog and digital transceiver architectures have to be considered with practical numbers of radio frequency (RF) chains\cite{TCOM22Cui,WCL22Chen,TWC23Shi}. In this regard, during the downlink training phase, the received signal at the UE from the BS in the $t$-th training slot can be expressed as
\begin{equation}
	{\bf y}^{(t)} = \left( {\bf W}_{\rm RF}^{(t)}{\bf W}_{\rm BB}^{(t)} \right)^{H}\left( {\bf H} {\bf F}_{\rm RF}^{(t)}{\bf F}_{\rm BB}^{(t)}{\bf s}^{(t)} + {\bf n}^{(t)} \right),
	\label{eq:dltrain}
\end{equation}
where ${\bf W}_{\rm RF}^{(t)}\in\mathbb{C}^{N_{\rm R}\times N^{\rm RF}_{\rm R}}$ and ${\bf W}_{\rm BB}^{(t)}\in\mathbb{C}^{N^{\rm RF}_{\rm R}\times N^{\rm S}_{\rm R}}$ denote the hybrid combiner matrices, whereas ${\bf F}_{\rm RF}^{(t)}\in\mathbb{C}^{N_{\rm T}\times N^{\rm RF}_{\rm T}}$ and ${\bf F}_{\rm BB}^{(t)}\in\mathbb{C}^{N^{\rm RF}_{\rm T}\times N^{\rm S}_{\rm T}}$ denote the hybrid precoders, respectively. $N_{\rm R}^{\rm RF}$ ($N_{\rm T}^{\rm RF}$) and $N_{\rm R}^{\rm S}$ ($N_{\rm T}^{\rm S}$) denote the numbers of RF chains and data streams at the receiver (transmitter), respectively. ${\bf n}^{(t)}\sim\mathcal{CN}(0,\sigma_{\rm n}^2 {\bf I})$ is the additive white Gaussian noise (AWGN) vector, and ${\bf s}^{(t)}$ denotes the pilot signal. 
	
From \eqref{eq:sv}, \eqref{eq:LoS}, and Fig.~\ref{fig:sysmodel}, it can be determined that each element in the near-field steering vector $\tilde{g}({\bf r}_{\rm T}^{(m)},{\bf r}_{\rm R}^{(n)}) = e^{-jk\sqrt{\Vert{\bf r}_{\rm T}^{(m)}\Vert^2+\Vert{\bf r}_{\rm R}^{(n)}\Vert^2-2\Vert{\bf r}_{\rm T}^{(m)}\Vert\Vert{\bf r}_{\rm R}^{(n)}\Vert\cos(\vartheta_{m,n})}}/\Vert{\bf r}_{\rm T}^{(m)}-{\bf r}_{\rm R}^{(n)}\Vert$ requires information in both the distance and angular domains. This is the main difference between the near-field channel model and the conventional far-field counterpart, where only angular information is decisive~\cite{TWC23Shi}. Hence, the inclusion of additional parameters related to distance introduces heightened complexity in CE problems.

\section{Problem Formulation}
\label{sec:formulation}
In this section, we exploit the sparsity of the near-field XL-MIMO channel and formulate the CE problem by capitalizing on the CS technique. Define ${\bf W}^{(t)} = ({\bf W}_{\rm RF}^{(t)}{\bf W}_{\rm BB}^{(t)})^H$ and ${\bf f}^{(t)} ={\bf F}_{\rm RF}^{(t)}{\bf F}_{\rm BB}^{(t)}{\bf s}^{(t)}$ for notational brevity, the signal model in \eqref{eq:dltrain} can be rewritten as ${\bf y}^{(t)} = \left( ({\bf f}^{(t)})^{T}\otimes {\bf W}^{(t)} \right){\rm vec}({\bf H})+\tilde{\bf n}^{(t)}$, where $\otimes$ denotes the Kronecker product, ${\rm vec}(\cdot)$ denotes the vectorization operation, and $\tilde{\bf n}^{(t)} = {\bf W}^{(t)}{\bf n}^{(t)}$. Stacking $\tau$ training slots together, we obtain
\begin{equation}
	{\bf y} = {\boldsymbol{\Phi}{\bf h}}+\tilde{\bf n},
	\label{eq:linearproblem}
\end{equation}
where ${\bf y} = [({\bf y}^{(1)})^{H},\cdots,({\bf y}^{(\tau)})^{H}]^{H}$ is the overall received signal, $\boldsymbol{\Phi} = [(({\bf f}^{(1)})^{T}\otimes {\bf W}^{(1)})^{H},\cdots, (({\bf f}^{(\tau)})^{T}\otimes {\bf W}^{(\tau)})^{H}]^{H}$ is the measurement matrix, and ${\bf h} = {\rm vec}({\bf H})$ is the vectorized downlink channel vector. Estimating $\bf h$ in \eqref{eq:linearproblem} via linear methods requires 
excessive training overhead $\tau\geq N_{\rm T}N_{\rm R}$, which is infeasible in XL-MIMO systems. In light of this, CS-based reconstruction methods were proposed to fully utilize the intrinsic sparsity of ${\bf H}$~\cite{TIT06Donoho}, and the sparse reconstruction problem can be formulated as
\begin{equation}
	\begin{aligned}
		{\rm(P1)}\quad\quad\underset{\tilde{\bf h}}{\min}\ &\ \Vert \tilde{\bf h} \Vert_0\\
		{\rm s.t.}\ &\ \Vert\boldsymbol{\Phi} \boldsymbol{\Psi} \tilde{\bf h}-\mathbf{y}\Vert_2 \leq \varepsilon,
	\end{aligned}
\end{equation}
where $\tilde{\bf h}$ is the sparse support vector to be estimated, $\varepsilon$ is the error bound, and ${\boldsymbol{\Psi}}$ is the codebook matrix. A desirable codebook should match the signal model of ${\bf h}$ to capture inherent features and efficiently sparsify the channel vector as $\tilde{\bf h}$. Besides, the mutual correlation between codewords in ${\boldsymbol{\Psi}}$ should be sufficiently low to avoid converging to multiple similar sparse representations that cause ambiguity~\cite{SPL17Miandji}.

In conventional far-field CE problems, the channel matrix can be efficiently sparsified by steering matrices with uniform angular domain sampling (i.e., DFT matrices) as ${\bf H} = {\bf A}_{\rm R}^{\rm D}\tilde{\bf H}({\bf A}_{\rm T}^{\rm D})^H$, where 
\begin{equation}
	\begin{aligned}
		{\bf A}_{{\rm R}}^{\rm D} &= \left[{\bf a}_{{\rm R}}(\theta_1),\cdots,{\bf a}_{{\rm R}}(\theta_{\beta N_{{\rm R}}})\right]\in\mathbb{C}^{N_{{\rm R}}\times \beta N_{{\rm R}}}.
	\end{aligned}
	\label{eq:codebookdft}
\end{equation}
${\bf a}_{{\rm R}}(\theta) = [1,e^{j\pi \sin\theta},\cdots,e^{j\pi(N_{{\rm R}}-1) \sin\theta}]^{H}$ in \eqref{eq:codebookdft} is the far-field steering vector, $\beta\geq 1$ is the oversampling rate, and $\theta_i = -\pi/2+i\pi/\beta N_{\rm R}$, where $i=1,\cdots,\beta N_{\rm R}$. Note that the steering matrix ${\bf A}_{{\rm T}}^{\rm D}$ at the transmitter side entails a similar form to ${\bf A}_{{\rm R}}^{\rm D}$. Given that ${\bf h}={\rm vec}({\bf A}_{\rm R}^{\rm D}\tilde{\bf H}({\bf A}_{\rm T}^{\rm D})^H) = (\left({\bf A}_{\mathrm{T}}^{\mathrm{D}}\right)^* \otimes {\bf A}_{\mathrm{R}}^{\mathrm{D}}) \tilde{\bf h} $, codebook $\boldsymbol{\Psi}$ is typically designed as $\boldsymbol{\Psi}=\left({\bf A}_{\mathrm{T}}^{\mathrm{D}}\right)^* \otimes {\bf A}_{\mathrm{R}}^{\mathrm{D}}\in\mathbb{C}^{N_{\rm R}N_{\rm T}\times \beta^2 N_{\rm R}N_{\rm T}}$.
	
However, the near-field channel matrix modeled in \eqref{eq:channelmodel} can no longer be properly sparsified by the far-field steering matrices in \eqref{eq:codebookdft} due to the {model mismatch between ${\mathbf g}_{\rm R}(\cdot)$ and ${\bf a}_{\rm R}(\cdot)$}, which will lead to a significant power leakage, increasing the number of iterations in CS-based CE algorithms, and degrading the channel reconstruction accuracy~\cite{TSP20Ke,TCOM22Cui}. 
	
\section{Proposed Channel Estimation based on {Eigenfunction} Representations}
	
In this section, we propose a novel codebook design to combat the challenges introduced by the model mismatch. Specifically, we employ EVD to the auto-correlation matrix of the near-field channel and derive the general form of the eigenvectors. The eigen-codebook is therefore constructed based on the eigenvectors, which are able to efficiently sparsify the near-field channel matrices. Furthermore, a two-step CE scheme is proposed to fully exploit the advantages of the proposed codebook. 
	
\subsection{Codebook Design}
Recall that problem $\rm (P1)$ requires the identification of a codebook $\boldsymbol{\Psi}$ that efficiently sparsifies the near-field channel matrix. In this regard, the singular value decomposition (SVD) decomposes the channel matrix in the form of ${\bf H} = {\bf U}{\boldsymbol{\Sigma}}{\bf V}^{H}$, where ${\bf H}$ can be properly sparsified to a diagonal singular value matrix ${\boldsymbol{\Sigma}}$ by unitary matrices ${\bf U}$ and ${\bf V}$. The resulting codebook $\boldsymbol{\Psi} = {\bf V}^*\otimes {\bf U}$ also shows mutual orthogonality between codewords. Inspired by the SVD, we consider designing the codebook matrix exploiting the singular vectors. Since the channel matrix is not a square matrix when $N_{\rm R}\neq N_{\rm T}$, singular vectors can be obtained separately from the corresponding EVD of the auto-correlation matrices. For the transmit eigenvectors, we first define the auto-correlation matrix by
\begin{equation}
	\begin{aligned}
		{\bf R}_{\rm T} & \overset{~~~}{=} \mathbb{E}\left[{\bf H}^{H} {\bf H}\right]\\
		& \overset{~~~}{=} \frac{K}{1+K}{\bf H}_{\rm LoS}^{H}{\bf H}_{\rm LoS}+\frac{1}{1+K}\mathbb{E}\left[ {\bf H}_{\rm NLoS}^{H}{\bf H}_{\rm NLoS} \right]\\
		&\overset{~~~}{=} \gamma K{\bf H}_{\rm LoS}^{H}{\bf H}_{\rm LoS}+\gamma{\bf I},
	\end{aligned}
	\label{eq:autocorr}
\end{equation}
where we denote $\gamma=1/(1+K)$ for notational brevity. The identity matrix on the right-hand side has no impact on calculating eigenvectors {since it only adds $\gamma$ to each eigenvalue}. Therefore, the element located at the $m^\prime$-th row and $m$-th column of ${\bf R}_{\rm T}$ can be expressed by
\begin{equation}
	\begin{aligned}
		{\bf R}_{\rm T}[m^\prime,m] &= \gamma K{\mathbf g}_{\rm R}^H({\bf r}_{\rm T}^{(m)}){\mathbf g}_{\rm R}({\bf r}_{\rm T}^{(m)})+\gamma\mathds{1}_{m,m^\prime}\\&=\gamma\mathds{1}_{m,m^\prime}+\gamma K\sum_{n=1}^{N_{\rm R}}\frac{e^{-j\kappa \Vert{\bf r}_{\rm T}^{(m)}-{\bf r}_{\rm R}^{(n)}\Vert }}{\Vert{\bf r}_{\rm T}^{(m)}-{\bf r}_{\rm R}^{(n)}\Vert} \\
		&\quad\times\frac{e^{j\kappa \Vert{\bf r}_{\rm T}^{(m^\prime)}-{\bf r}_{\rm R}^{(n)}\Vert}}{\Vert{\bf r}_{\rm T}^{(m^\prime)}-{\bf r}_{\rm R}^{(n)}\Vert},
	\end{aligned}
	\label{eq:autocorr2}
\end{equation}
where $\mathds{1}_{m,m^\prime}$ is the indicator function. Introducing the near-field paraxial approximation\cite{AO20Miller}, we have
\begin{align}
	{\bf R}_{\rm T}[m^\prime,m] \approx &~ \gamma\mathds{1}_{m,m^\prime}\notag\\&~ +\frac{\gamma K}{r_0^2}\sum_{n=1}^{N_{\rm R}}e^{-j\kappa\frac{\left(x_{\rm T}^{(m)}-x_{\rm R}^{(n)}\right)^2-\left(x_{\rm T}^{(m^\prime)}-x_{\rm R}^{(n)}\right)^2}{2y_0}}\notag\\
	=&~\gamma\mathds{1}_{m,m^\prime}+\frac{\gamma K e^{j\kappa \frac{(x_{\rm T}^{(m^\prime)})^2-(x_{\rm T}^{(m)})^2}{2y_0}}}{r_0^2 } \label{eq:dft}\\
	&~\times\sum_{n=1}^{N_{\rm R}} e^{j\kappa\frac{x_{\rm R}^{(n)}\left(x_{\rm T}^{(m)}-x_{\rm T}^{(m^\prime)}\right)}{y_0} }\notag\\
	\triangleq&~\gamma\mathds{1}_{m,m^\prime}+\gamma K e^{j\kappa \frac{(x_{\rm T}^{(m^\prime)})^2-(x_{\rm T}^{(m)})^2}{2y_0}} {\bf R}^\prime_{\rm T}[m^\prime,m],\notag
\end{align}
where $r_0$ is an approximation of the distance term in the denominator of the second term in \eqref{eq:autocorr2} and $y_0$ denotes the center of the $y$-coordinate of the UE array. Hence, we can rewrite the EVD procedure of ${\bf R}_{\rm T}$ as 
\begin{equation}
	{\bf R}_{\rm T}{\bf v}_m = {\bf D}_{\rm T}^{-1} \left( \gamma K\mathbf{R}^\prime_{\rm T}+\gamma {\bf I} \right){\bf D}_{\rm T}{\bf v}_m = \lambda_m{\bf v}_m,
	\label{eq:evd}
\end{equation}
where ${\bf v}_m$ is the $m$-th eigenvector of ${\bf R}_{\rm T}$ and $\lambda_m$ is the corresponding eigenvalue. The compensation matrix ${\bf D}_{\rm T}$ is an $(x_{\rm T}^{(m)})^2$-related phase term extracted from ${\bf R}_{\rm T}$ according to~\eqref{eq:dft} as
\begin{equation}
	{\bf D}_{\rm T} = {\rm diag}(e^{j\kappa \frac{(x_{\rm T}^{(1)})^2}{2y_0}},\cdots,e^{j\kappa \frac{(x_{\rm T}^{(N_{\rm T})})^2}{2y_0}}).
	\label{eq:compensation}
	\vspace{-2mm}
\end{equation}

Note that extracting ${\bf D}_{\rm T}$ from ${\bf R}_{\rm T}$ only changes the phase of each eigenvector since $\left( \gamma K\mathbf{R}^\prime_{\rm T}+\gamma {\bf I} \right){\bf D}_{\rm T}{\bf v}_m = \lambda_m {\bf D}_{\rm T} {\bf v}_m$. Furthermore, \eqref{eq:dft} yields
\begin{equation}
	\begin{aligned}
		{\bf R}_{\rm T}^\prime[m^\prime,m] & \overset{~~~}{=} \frac{1}{r_0^2} \sum_{n=1}^{N_{\rm R}} e^{j\kappa\frac{x_{\rm R}^{(n)}\left(x_{\rm T}^{(m)}-x_{\rm T}^{(m^\prime)}\right)}{y_0} }\\
		&\overset{(a)}{\approx} \frac{1}{r_0^2} \int_{-L_{\rm R}/2}^{L_{\rm R}/2}e^{\frac{j\kappa}{y_0}x (x_{\rm T}^{(m)}-x_{\rm T}^{(m^\prime)})} {\rm d}x \\&\overset{~~~}{=}\frac{ 2y_0\sin\left[ \frac{\kappa L_{\rm R} (x_{\rm T}^{(m)}-x_{\rm T}^{(m^\prime)} )}{2y_0} \right]  }{ r_0^2 \kappa \left(x_{\rm T}^{(m)}-x_{\rm T}^{(m^\prime)} \right)}\\ &\overset{~~~}{\propto}\frac{\sin\left[2\pi W (x_{\rm T}^{(m)}-x_{\rm T}^{(m^\prime)} \right]  }{ \left(x_{\rm T}^{(m)}-x_T^{(m^\prime)} \right)},
	\end{aligned}
	\label{eq:toeplitzmat}
\end{equation}
where $L_{\rm R}=(N_{\rm R}-1)\lambda/2$ denotes the aperture of the UE array with half-wavelength antenna spacing. Note that $(a)$ asymptotically holds when $N_{\rm R}$ is sufficiently large. ${\bf R}_{\rm T}^\prime$ is a Toeplitz matrix with each column composed of a \textit{shifted sinc function}, and the $m$-th eigenvector ${\bf D}_{\rm T} {\bf v}_m$ of this matrix is called the $(m-1)$-th order \emph{discrete prolate spheroidal sequence} (or Slepian sequence) within frequency $W=\kappa L_{\rm R}/(4\pi y_0)$~\cite{Slepian54TIT}.
\begin{remark}
	Typically, estimating the auto-correlation matrix requires a large number of samples. However, with the result in~\eqref{eq:toeplitzmat}, the auto-correlation matrix can be well-determined directly by a series of sinc functions given the frequency $W$, from which we can generate the codebook using an efficient EVD operation.
\end{remark}

Similarly, we can calculate the eigenvectors $\{ {\bf u}_n \}_{n=1}^{N_{\rm R}}$ of ${\bf R}_{\rm R} = {\bf H}{\bf H}^{H}=\mathbf{U}\boldsymbol{\Lambda}^\prime\mathbf{U}^{H}\in\mathbb{C}^{N_{\rm R}\times N_{\rm R}}$ and finally form the eigen-codebook matrix as ${\boldsymbol{\Psi}}_e = \mathbf{D_{\rm T}V}^* \otimes \mathbf{D_{\rm R}U} \in\mathbb{C}^{N_{\rm R}N_{\rm T}\times N_{\rm R}N_{\rm T}}$ for problem $\rm (P1)$. By resorting to the EVD tailored for the near-field channel matrix, we can effectively eliminate the mismatch issue associated with the DFT codebook. Moreover, the proposed DPSS-based eigen-codebook naturally holds orthogonality among columns, since both $\bf V$ and $\bf U$ are unitary matrices. This is one of the key advantages compared to the spherical codebook~\cite{TCOM22Cui}, which shall be validated via simulation in the next section.

\subsection{Proposed Two-Step Near-Field Channel Estimation}
\label{sec:proposed}
To calculate the near-field eigen-codebook matrix, we need to know the (approximate) location of the UE\footnote{Establishing a coordinate system with the BS as the origin can avoid dependence on its location information, but we still require the location of the UE.}. In this subsection, we propose a two-step algorithm for near-field CE, which firstly estimates the location and then designs the eigen-codebook to solve the sparse reconstruction problem $\rm (P1)$. The two-step CE procedure can be given as
\begin{enumerate}
	\item \textbf{Coarse Localization}: Construct a spherical wave codebook ${\boldsymbol{\Psi}}_p$ with angle and distance sampled in the polar-domain~\cite{TCOM22Cui} for coarse location estimation as
	\begin{equation}
		i={\rm arg}\underset{j}{\rm max}\ \left\Vert \left( {\boldsymbol{\Psi}}_p^{H} \right)[j,:] {\boldsymbol{\Phi}}^{H} {\bf y} \right\Vert^2,
		\label{eq:locest}
	\end{equation}
	from which the location coordinates $({\hat x}_i,{\hat y}_i)$ can be obtained through index-coordinate mapping of the codebook ${\boldsymbol{\Psi}}_p$.
	\item {\bf Channel Estimation with the proposed DPSS-based Eigen-Codebook}:
	Calculate the compensation matrix $\hat{\bf D}_{\rm T}$ (or $\hat{\bf D}_{\rm R}$) in \eqref{eq:evd} with estimated coordinates\footnote{According to the paraxial approximation in \eqref{eq:dft}, employing $\exp ( {-j\kappa\frac{(\hat{x}_i)^2}{2\hat{y}_i}} )$ for all diagonal elements to construct $\hat{\bf D}_{\rm T}$ (or $\hat{\bf D}_{\rm R}$) provides an accurate approximation of ${\bf D}_{\rm T}$ (or ${\bf D}_{\rm R}$) as expressed in \eqref{eq:evd}.\label{fn}} $({\hat x}_i,{\hat y}_i)$. Construct the DPSS-based eigen-codebook ${\boldsymbol{\Psi}}_e$ according to \textbf{Algorithm~\ref{alg:a1}} and employ CE with CS-based algorithms such as orthogonal matching pursuit (OMP)~\cite{TWC18Rod}.
\end{enumerate}

\SetAlFnt{\small}
\SetAlCapFnt{\normalsize}
\SetAlCapNameFnt{\normalsize}
\begin{algorithm}[!t]
	\caption{Proposed Codebook Design Algorithm}\label{alg:a1}
	\begin{algorithmic}[1]
		\REQUIRE Estimated coordinate $(\hat{x}_i,\hat{y}_i)$ and the numbers of \\antennas $N_{\rm T}$ and $N_{\rm R}$.
		\ENSURE The DPSS-based eigen-codebook $\boldsymbol{\Psi}_e$.
		\STATE Estimate the compensation matrix $\hat{\bf D}_{\rm T}$ and $\hat{\bf D}_{\rm R}$ according to~\eqref{eq:compensation}.
		\STATE Calculate the frequency ${\hat W}=\kappa L_{\rm R}/(4\pi \hat{y}_0)$.
		\STATE Generate ${\bf R}_{\rm T}$ and ${\bf R}_{\rm R}$ with DPSS according to \eqref{eq:toeplitzmat}.
		\STATE Perform EVD for ${\bf R}_{\rm T}=\mathbf{V}\boldsymbol{\Lambda}\mathbf{V}^{-1}$ and ${\bf R}_{\rm R}=\mathbf{U}{\boldsymbol{\Lambda}^\prime}\mathbf{U}^{-1}$.
		\STATE Compensate phase shift ${\bf V}^c=\hat{\bf D}_{\rm T} {\bf V}$, ${\bf U}^c=\hat{\bf D}_{\rm R} {\bf U}$.
		\STATE Return eigen-codebook ${\boldsymbol{\Psi}}_e = ({\bf V}^c)^* \otimes {\bf U}^c$.
	\end{algorithmic}
\end{algorithm}

\section{Simulation Results}
In this section, we evaluate the channel reconstruction performance based on the proposed eigen-codebook via numerical simulations. The performance is evaluated by normalized mean square error (NMSE) as
\begin{equation}
	{\rm NMSE}\left( \hat{\bf H} ,{\bf H} \right) = \mathbb{E}\left[{\Vert \hat{\bf H} -{\bf H} \Vert_F^2}/{\Vert {\bf H} \Vert_F^2}\right],
\end{equation}
where $\Vert\cdot\Vert_F$ is the Frobenius norm, and $\hat{\bf H}$ is an estimation of ${\bf H}$.
\subsection{Simulation Setup}
\label{sec:setup}
Throughout the simulation, we consider the large arrays at the BS and UE are equipped with $N_{\rm T}=192$ and $N_{\rm R} = 4$ antennas with half-wavelength spacing, respectively, and the carrier frequency is set as $f_c = 28\,{\rm GHz}$. The BS array is placed symmetrically on  the $x$-axis and the UE is in the near-field region of the BS as shown in Fig.~\ref{fig:sysmodel}. Unless otherwise specified, we deploy a single RF chain at both the BS and UE. The distance from the UE to the center of the BS array is selected uniformly from $[1\,{\rm m},\:20\,{\rm m}]$, and the Rician factor is set to $K = 13\:{\rm dB}$~\cite{TWC18Rod}.

We mainly consider three types of codebooks in the simulation, namely the DFT codebook~\cite{TWC18Rod}, the spherical wave codebook in the polar-domain~\cite{TCOM22Cui}, and the proposed DPSS codebook. For the DFT codebook, we set the number of angle grids as $\beta N_{\rm T}$ and $\beta N_{\rm R}$ at the BS and UE, respectively, with $\beta$ being the oversampling rate in~\eqref{eq:codebookdft}. For the spherical wave codebook, both angle and distance grids are set as $\nint{\beta \sqrt{N_{\rm T}}}$ and $\nint{\beta \sqrt{N_{\rm R}}}$ at the BS and UE, respectively, where $\nint{\cdot}$ denotes the rounding operator. In this case, as was mentioned in Section~\ref{sec:formulation}, the sizes of the DFT codebook and spherical wave codebook are $\beta^2N_{\rm T}N_{\rm R}$ and $\nint{\beta \sqrt{N_{\rm R}}}\nint{\beta \sqrt{N_{\rm T}}}\simeq \beta^2N_{\rm T}N_{\rm R}$, respectively. Note that the size of the proposed DPSS-based codebook is irrelevant to $\beta$ because the number of eigenvectors will not change. Additionally, the compressive ratio (CR) of sparse reconstruction problem is defined as $\mu = \tau/(N_{\rm R}N_{\rm T})$. For fair comparison, the performance achieved by all codebooks is evaluated based on the OMP algorithm.

\subsection{Numerical Results}
\label{sec:results}
\begin{figure}[t]
	\centering
	\includegraphics[width=0.4\textwidth]{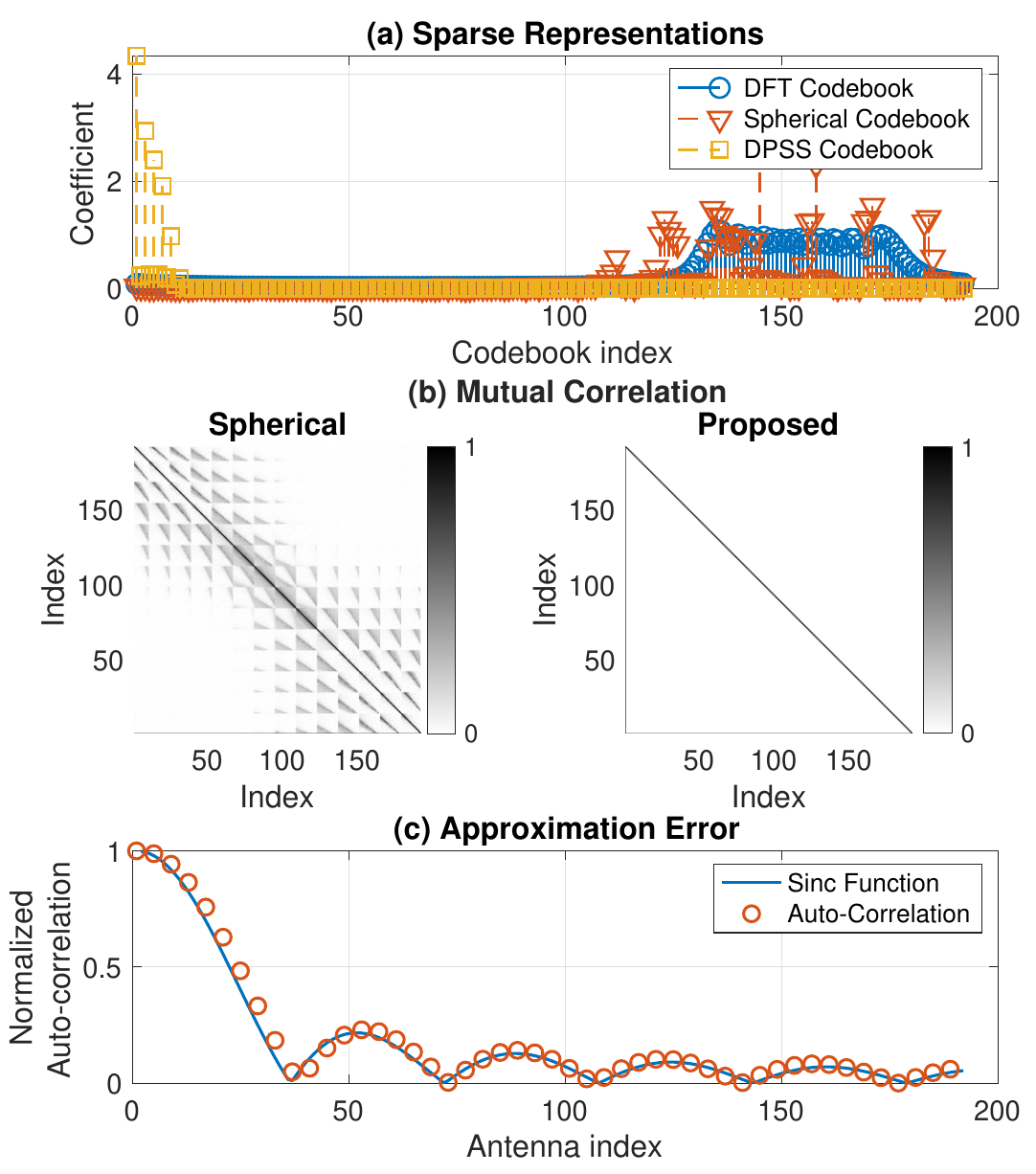}
	\caption{(a) The sparse representations of near-field channels under different codebooks, (b) the mutual correlation matrix $\boldsymbol{\Psi}^H\boldsymbol{\Psi}$ of the spherical wave codebook and the proposed codebook, and (c) the approximation error in~\eqref{eq:toeplitzmat}.}
	\label{fig:res1}
	\vspace{-3mm}
\end{figure}

We firstly investigate the sparsification performance of the proposed codebook. The channel sparse representations of the DFT codebook, spherical wave codebook, and proposed codebook are compared in Fig.~\ref{fig:res1}(a), where the sparse representation is obtained by $\tilde{\bf h} = \boldsymbol{\Psi}^\dagger {\bf h}$. As can be observed, the conventional DFT codebook shows a severe energy leakage effect in the near-field region, which can be improved by the spherical wave codebook sampled in the polar-domain. Meanwhile, the proposed DPSS-based eigen-codebook entails the sparsest pattern among the three codebooks. Note that the proposed codebook is compensated by the matrix $\hat{\bf D}_{{\rm T}({\rm R})}$ and therefore, the sparse representation shows no specific angular information. In addition, different from the DFT and spherical codebooks, the support appears in the first several indices since the SVD always sorts the non-zero singular values first.

On the other hand, Fig.~\ref{fig:res1}(b) plots the colormap that represents the values of $\boldsymbol{\Psi}^H\boldsymbol{\Psi}$. As can be observed, the codewords in the proposed DPSS-based codebook are strictly orthogonal to each other, which is far beyond the capabilities of the spherical codebook. We further validate the approximation error of the derivation procedure in \eqref{eq:toeplitzmat}. As is depicted in Fig.~\ref{fig:res1}(c), the auto-correlation curve stands for the absolute value of ${\bf R}_{\rm T}^\prime[1,:]$, while the red circles show the value of the normalized sinc function. The approximation procedure shows negligible error, which confirms the high accuracy of our proposed approximation in~\eqref{eq:toeplitzmat}.

\begin{figure}[t]
	\centering
	\includegraphics[width=0.4\textwidth]{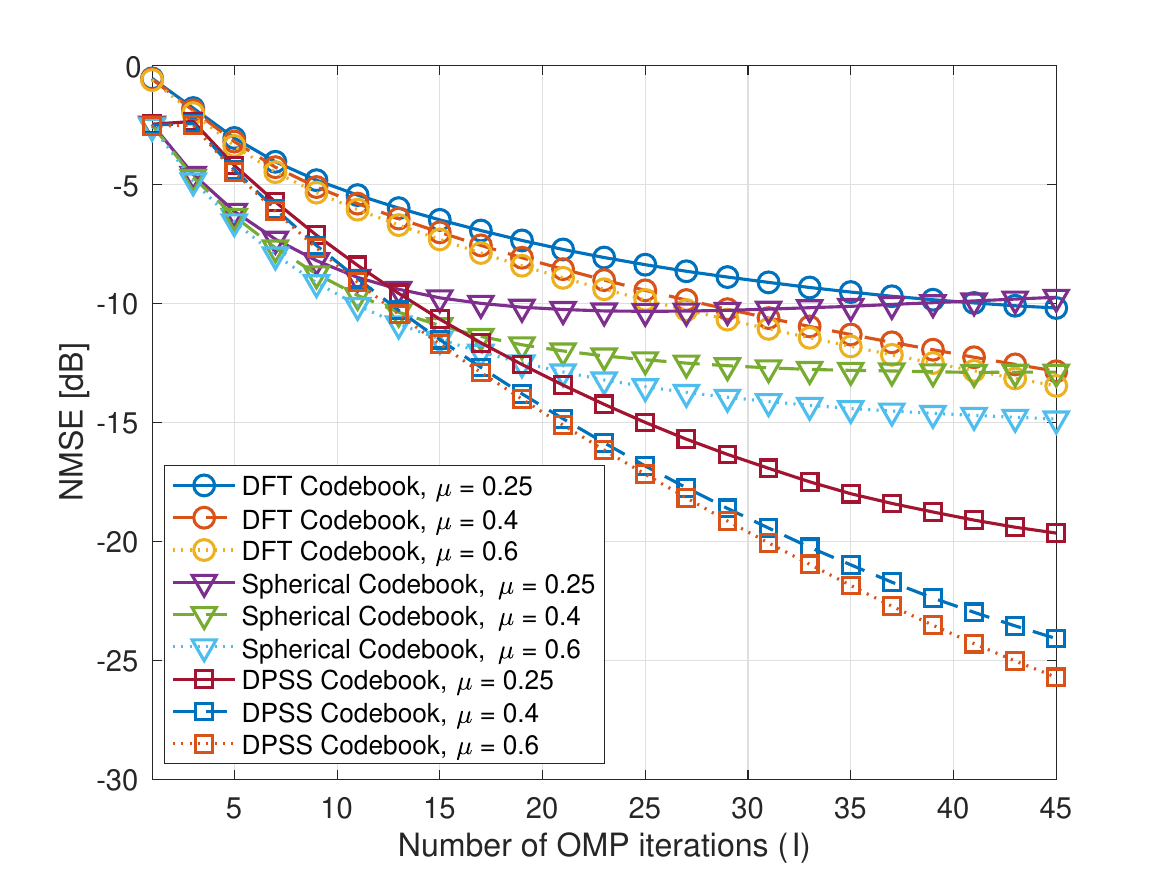}
	\caption{Performance comparison of near-field CE error versus the CR $\mu$ and iterations $I$.}
	\label{fig:res2}
	\vspace{-2mm}
\end{figure}

We then investigate the CE accuracy performance with CR $\mu = \{0.25, 0.4, 0.6\}$. The oversampling rate $\beta$ is set to $1$ to keep the sizes of the three considered codebooks identical. The reconstruction accuracy increases as $\mu$ increases, where the proposed DPSS-based eigen-codebook achieves the best NMSE performance as shown in Fig.~\ref{fig:res2}. In particular, at $I=1$ the proposed method shows the same performance as the spherical wave method since we regard the first-step coarse localization in Section~\ref{sec:proposed} as one iteration. The slight performance drop at $I=2$ is also due to an abrupt codebook switch at the second step of the proposed CE scheme. Starting from $I=10$, thanks to the excellent ability to sparsify the near-field channel with mutually orthogonal codewords, the proposed DPSS-based eigen-codebook outperforms the baselines by a large margin and converges to the lowest NMSE among the considered codebooks.

\begin{figure}[t]
	\centering
	\includegraphics[width=0.4\textwidth]{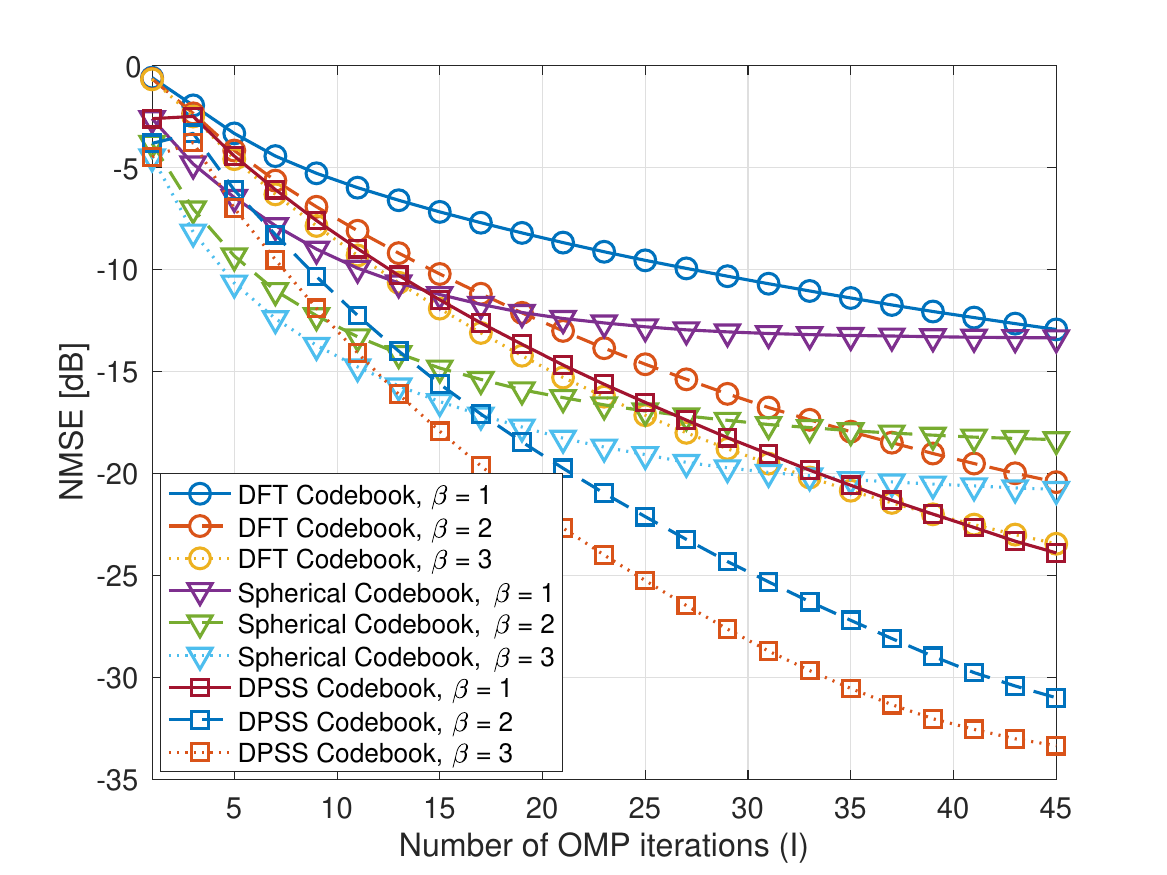}
	\caption{Performance comparison of near-field CE error with CR $\mu = 0.4$ versus the number of iterations $I$ and codebook oversampling rate $\beta$.}
	\label{fig:res3}
	\vspace{-2mm}
\end{figure}

We then evaluate the CE performance with oversampling rates $\beta=\{1,2,3\}$ when CR $\mu=0.4$. As is shown in Fig.~\ref{fig:res3}, We can see significant performance improvement for all schemes by increasing $\beta$, while the proposed DPSS-based codebook still achieves the highest reconstruction accuracy within sufficient iterations. However, the performance gains for the DFT and spherical codebooks are achieved at the cost of larger codebook sizes. Specifically, as mentioned in Section~\ref{sec:setup}, their sizes increase quadratically with $\beta$, i.e., $\beta^2N_{\rm T}N_{\rm R}$. In contrast, the increase in $\beta$ only affects the localization accuracy in the first step of our proposed CE scheme while the size of the DPSS-based codebook remains $N_{\rm T}N_{\rm R}$. In other words, the CE performance achieved by the DPSS-based codebook tremendously outperforms those of two baselines even with a much smaller codebook size.

\begin{figure}[t]
	\centering
	\includegraphics[width=0.4\textwidth]{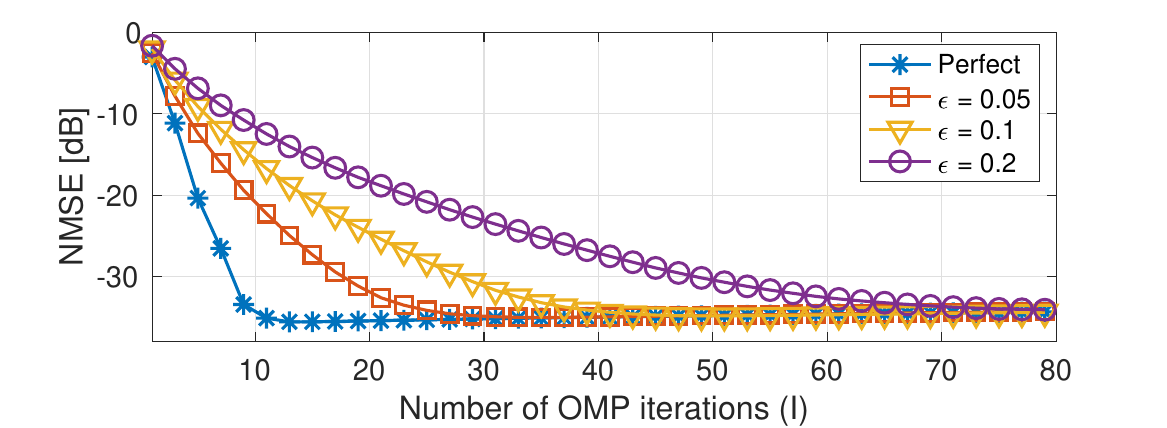}
	\caption{Performance of proposed DPSS-based eigen-codebook versus different localization errors $\epsilon$.}
	\label{fig:res4}
\end{figure}

As mentioned in Section~\ref{sec:proposed}, the proposed CE scheme involves a coarse localization as the first step. In Fig.~\ref{fig:res4}, we investigate the NMSE performance versus different localization errors $\epsilon = \sqrt{(\hat{x}_i-\bar{x}_{\rm R})^2+(\hat{y}_i-\bar{y}_{\rm R})^2}~({\rm m})$, which denotes the distance from the center of the UE array $(\bar{x}_{\rm R},\bar{y}_{\rm R})$ to the estimated location coordinate $({\hat x}_i,{\hat y}_i)$ in \eqref{eq:locest}. $\epsilon$ is assumed to be uniformly distributed within a circular area. As can be observed, the proposed codebook ensures the convergence of the OMP algorithm within the considered error levels\footnote{According to the recent field-test~\cite{TWC22Sakhnini}, the $95$th percentile of the localization error is observed to be around $0.2\:{\rm m}$.}, while more OMP iterations are required for a larger value of $\epsilon$. This result demonstrates the robustness of the proposed CE scheme against the localization error.

\subsection{Storage Analysis}

We further evaluate the storage requirements of the codebook given a target convergence NMSE. As is shown in Table~\ref{tab:1}, we compare the minimum required codebook size, i.e., the number of codewords, to achieve the NMSE targets $\{-15, -20, -25, -30\}~{\rm dB}$. The sizes of the DFT and spherical codebooks keep increasing with higher NMSE requirements, while the DPSS-based codebook size remains constant. Additionally, the $-30$ $\rm dB$ NMSE cannot be achieved by enlarging the sizes of the two baseline codebooks, and the corresponding sizes are displayed as ${\rm N/A}$. In particular, thanks to its mutual orthogonality among codewords, the DFT codebook can satisfy more stringent NMSE requirements with only slightly larger sizes. Yet, its mismatch with the near-field channel model still leads to a bulkier codebook compared to the proposed DPSS-based one. On the other hand, the two DoFs in both distance and angle of the spherical wave codebook dramatically add to the codebook size as the resolution requirement increases. Compared to the DFT and spherical wave codebook, the proposed DPSS-based codebook does not need to sacrifice NMSE performance for a lower storage, and its orthogonality enables it to converge faster than the spherical wave codebook.

\begin{table}[t]
	\centering
	\caption{The minimum required codebook size for target NMSE.}\label{tab:1}
	\begin{tabular}{c|cccc}
		\hline
		\hline
		\multirow{2}{*}{Codebooks} & \multicolumn{4}{c}{Target NMSE}                                      \\ \cline{2-5} 
		& \multicolumn{1}{l}{$-15\:$dB} & \multicolumn{1}{l}{$-20\:$dB} & \multicolumn{1}{l}{$-25\:$dB} & $-30\:$dB \\ \hline
		DFT                        &  $768$                          &  $855$                            &  $1,150$  &  $\rm N/A$   \\ \hline
		Spherical\cite{TCOM22Cui}                        &   $ 1,150$                           &  $3,072$                            &   $15,552$   & $\rm N/A$  \\ \hline
		Proposed                       &   $\bf 768$               &  $\bf 768$                 &  $\bf 768$   &  $\bf 768$ \\ \hline\hline
	\end{tabular}
	\vspace{-3mm}
\end{table}

\section{Conclusion}
In this paper, we proposed a novel DPSS-based eigen-codebook for near-field XL-MIMO CE. By leveraging the EVD associated with the near-field channel, the proposed codebook achieves mutual orthogonality among codewords, and outperforms conventional DFT and polar-domain spherical wave codebooks in channel sparsification. We further proposed a two-step CE scheme, with which our proposed DPSS-based codebook achieves the best NMSE performance in CE. Furthermore, we compared the minimum required codebook size for different NMSE targets, which proved the proposed codebook effectively reduces the storage requirements. 


%

\end{document}